\documentclass[fleqn,usenatbib]{mnras}

\usepackage{newtxtext,newtxmath}

\usepackage[T1]{fontenc}

\DeclareRobustCommand{\VAN}[3]{#2}
\let\VANthebibliography\thebibliography
\def\thebibliography{\DeclareRobustCommand{\VAN}[3]{##3}\VANthebibliography}

\usepackage{graphicx}	
\graphicspath{ {./images/} }
\usepackage{amsmath}	

\title[Probabilities of quasi-molecular recombination]{Influence of a quasi-molecular mechanism of recombination on the formation of hydrogen in the early universe}

\author[T. Kereselidze, I. Noselidze and J. F. Ogilvie]{
Tamaz Kereselidze$^{1}$\thanks{E-mail: tamaz.kereselidze@tsu.ge}, Irakli Noselidze$^{2}$ and John F. Ogilvie$^{3,4}$
\\
$^{1}$Faculty of Exact and Natural Sciences, Tbilisi State University, Chavchavadze Avenue 3, 0179 Tbilisi, Georgia\\
$^{2}$School of Science and Technology, University of Georgia, Kostava Str. 77a, 0171 Tbilisi, Georgia\\
$^{3}$Centre for Experimental and Constructive Mathematics, Department of Mathematics, Simon Fraser University, 8888 University Drive, Burnaby, British\\
Columbia V5A 1S6, Canada\\
$^{4}$Escuela de Quimica, Universidad de Costa Rica, Ciudad Universitaria Rodrigo Facio, San Pedro de Montes de Oca, San Jose 11501-2060 Costa Rica
}

\date{Accepted XXX. Received YYY; in original form ZZZ}

\pubyear{2020}

\begin{document}
\label{firstpage}
\pagerange{\pageref{firstpage}--\pageref{lastpage}}
\maketitle

\begin{abstract}
In the framework of a quasi-molecular approach, the formation of hydrogen atom in the pre-recombination period of evolution of the universe is analysed quantitatively. Calculations in an adiabatic multi-level representation enable estimates of probabilities of radiative transitions. The quasi-molecular mechanism of recombination allows the formation of hydrogen molecular ion,  $H_2^+$, in its ground state. The probability of this process is comparable with the probability of the creation of atomic hydrogen. The participation of a second proton in the recombination increases the binding energy of an electron and decreases the rate of recombination of hydrogen.
\end{abstract}




\section{Introduction}

Cosmological recombination was responsible for the formation of neutral hydrogen and helium atoms in the early universe. For an electron and a proton the cosmological recombination was first studied by  \cite{Zeld68} and slightly latter by \cite {Peebles68}. Despite substantial progress achieved after these pioneering works there remain problems in understanding how the details of recombination affect the cosmological parameters. To explore this problem \cite{Liu2019} and \cite{ChSl} varied physical and phenomenological parameters in a standard code to compute the recombination history of the universe. They found that a cosmological parameter, the Hubble constant, is robust against perturbations of recombination history, unless non-standard physics modifies the atomic constants during the recombination epoch.

In our recent paper \citep{KerMNRAS} a quasi-molecular mechanism of recombination (QMR) was suggested and applied to treat the formation of atomic hydrogen in the early universe. According to this QMR, in the pre-recombination period of evolution of the universe $(z\gtrsim2000)$, when the temperature and density of protons were higher than subsequently, the recombination of an electron and a proton occurred in the presence of the nearest neighbouring proton, which participated in the process. An electron and two protons were considered to constitute quasi-molecule $H_2^+$ temporarily formed during a collision.

As an electron is much lighter than a proton, the velocity of an electron substantially exceeds a velocity of a proton in the quasi-molecule. This fact allows us to treat $H_2^+$  on a basis of an adiabatic representation. In this approximation all characteristics of  $H_2^+$, such as the electron binding energy, dipole strengths, quasi-molecular energy terms, profiles of spectral lines etc. depend upon the distance $R$  between protons.

According to the QMR a free electron emits a photon and creates $H_2^+$  in a highly excited state. Free-bound radiative transitions occur at distances between protons greater than the radius of the hydrogen atom in a highly excited state. If $H_2^+$  is formed in a repulsive state, the system rapidly dissociates into an excited hydrogen atom and a proton. The duration of dissociation is defined by the collision period, which is about $10^{-11}$ s for highly excited states and decreases to $10^{-14}$ s for the lowest states. From an excited state  $H$ descends to the state with principal quantum number $n=2$. A radiative decay from state $2{}^2P$  involving one photon or from state $2{}^2 S$  involving two photons then yields the hydrogen atom in its ground state.

If a quasi-molecule is formed in an attractive state, which can bind the colliding particles, a direct formation of the hydrogen atom is impossible. In this case radiative transitions lead to a cascade downward to low-lying attractive or repulsive quasi-molecular states. The QMR thus leads to a radiative transition of two types: free-bound with a direct formation of the hydrogen atom in the highly excited state, and free-bound with subsequent intermediate bound-bound quasi-molecular transitions that end with the formation of  $H$.

The main conclusion made by \cite{KerMNRAS} was that the radiative transition of an electron to an excited attractive state of $H_2^+$ affects the probability of recombination; the QMR should hence be included in a calculation of the cosmological recombination radiation.

The purpose of the present paper is to describe quantitatively the non-standard quasi-molecular mechanism of recombination. For this purpose, we implemented the appropriate calculations and answer this question: is the QMR significant for a complete study of the cosmological recombination problem? The treatment is performed in an adiabatic multi-level representation.

The paper is organized as follows. After stating our objective, we analyse the behaviour of the energy terms of $H_2^+$ in Sections 2, and evaluate radiative transition probabilities in Section 3. Using the obtained equations, we perform the appropriate calculations in Section 4, before a conclusion in Section 5. Unless otherwise indicated, atomic units  ($e=m_e=\hbar=1$) are used throughout the paper.

\section{Behavior of quasi-molecular energy terms}

 For our purpose it is important to know the behaviour of the energy terms of  $H_2^+$ at  large distances $R$ between protons. More precisely, for the QMR the existence of energy terms that are attractive is crucial, so that, accordingly, the colliding particles can bind during a period greater than a collision interval.

 At large $R$ the energy terms of $H_2^+$  are representable as \citep{BR1968}
\begin{equation}
U^{g,u}_{n_1,n_2,|m|}(R) =- \frac{1}{2n^2}+\frac{3n(n_1-n_2)}{2R^2}+O(R^{-3})\mp \Delta_{n_1,n_2,|m|}(R).
\label{eq_E1}
\end{equation}
Here the first three terms define the long-range interaction between the hydrogen atom and proton; the last term describes the exponentially small exchange interaction between the particles and is defined as \citep{KSP}
\begin{equation}
\begin{array}{l}
{\Delta _{{n_1}{n_2}\left| m \right|}}(R) = \frac{{{{( - 1)}^{\left| m \right|}}}}{{{n^3}{n_2}!({n_2} + \left| m \right|)!}}{\left( {\frac{{2R}}{n}} \right)^{n - {n_1} + {n_2}}}\\
 \cdot {e^{ - \frac{R}{n} - n}}\left( {1 + O({R^{ - 1}})} \right).
\end{array}
\label{eq_E2}
\end{equation}

In (\ref {eq_E1}) and (\ref {eq_E2}) $n_1,n_2,m$  are parabolic quantum numbers that specify electron states in the separate hydrogen atom; total quantum number $n$  is related to  $n_1,n_2,m$ with equation   $n=n_1+n_2+|m|+1$. Quasi-molecular energy terms $U^{g,u}_{n_1,n_2,|m|}(R)$ are distinguished by  parity, which is even (\emph{gerade}) or odd (\emph{ungerade}). As  is clear from equation (\ref {eq_E1}) the energy term is attractive at large $R$ if $n_2>n_1$. Among the terms with $n_1=n_2$ and $m=0$, the $g$ term is attractive and the $u$  term is repulsive.

\begin{figure}
\includegraphics[scale=0.8]{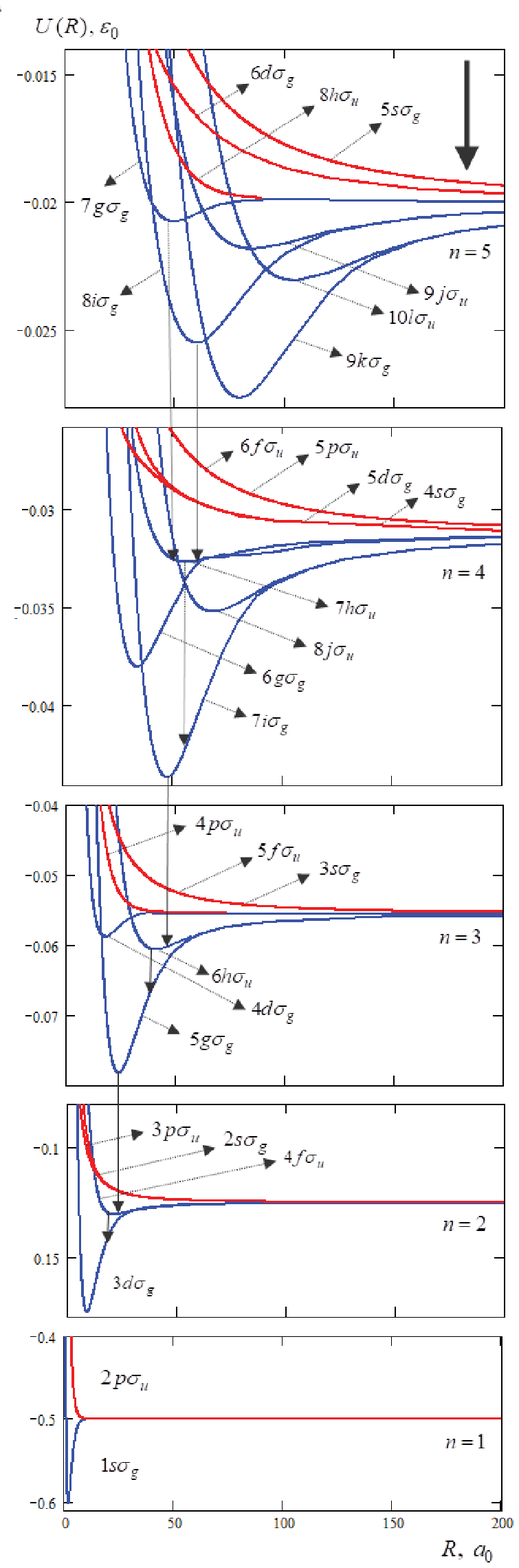}
\caption{
 $\sigma$ energy terms of $H_2^+$  as functions of  distance $R$
                       between protons, with blue curves for attractive terms and red curves for repulsive terms. Thin arrows indicate radiative transitions between
                       attractive quasi-molecular states;  ${a_0} = {\hbar ^2}/{m_e}{e^2} = 0.529 \times {10^{ - 8}}$cm
                          is the first Bohr radius of hydrogen and  ${\varepsilon _0} = {m_e}{e^4}/{\hbar ^2}=27.21$eV.
                                    Energy terms $(6f\sigma_u, 5p\sigma_u)$, $(5f\sigma_u,3s\sigma_g)$ and $( 3p\sigma_u,2s\sigma_g)$  are so close that they are indistinguishable in the figure.}
\label{fig:Figure 1}
\end{figure}

We proceed to investigate the behaviour of the energy terms of $H_2^+$ in the entire region of internuclear distances $R$. To avoid cumbersome calculations and at the same time to maintain generality, we restrict the treatment to the lowest thirty quasi-molecular terms with  $m=0$  ($\sigma$ terms). These terms correlate with the levels of the hydrogen atom with $n=1,2,3,4,5$ at $R=\infty$. The behaviour of the energy terms is depicted in Fig. ~\ref{fig:Figure 1}.

The quasi-molecular terms are specified with quantum numbers $n_0$  and $l_0$  that, together with $m$,  characterize an electron in the \emph{united atom} ($R=0$). Parabolic quantum numbers $n_1,n_2$  are related to quantum numbers $n_0, l_0, m$  according to the molecular-orbital correlation rules  $n_1=n_0-l_0-1$ and $n_2=\left(l_0-|m|\right)/2$  for $g$  orbitals and  $n_2=\left(l_0-|m|-1\right)/2$ for $u$  orbitals  \citep {BR1968, Ker1987}.

\section{Radiative transition probabilities}
In this Section, we evaluate the probabilities for radiative transitions involved in the QMR. Fig.~\ref{fig:Figure 1} shows that, among the states under consideration, the attractive ones are
\begin{equation}
 \begin{array}{l}
 1s, 3d, 4f, 5g, 6h, 4d, 7i, 8j,\\
6g, 7h,9k, 10l, 8i,9j, 7g,
\end{array}
\label{eq_E3}
\end{equation}
whereas the repulsive states are
\begin{equation}
 \begin{array}{l}
 2p, 2s, 3p, 5f, 3s, 4p, 5d, 6f,\\
4s, 5p,8h, 6d, 7f,5s, 6p.
\end{array}
\label{eq_E4}
\end{equation}
In (\ref {eq_E3}) and (\ref {eq_E4}) a symbol $\sigma$  is omitted and states are arranged in order of increasing  energy at large $R$; subscripts $g$  and $u$  are also omitted because the parity of a state is uniquely defined with quantum number $l_0$.

Adjusting a harmonic oscillator potential to the attractive energy term, one can  readily show that each potential well depicted in Fig.~\ref{fig:Figure 1} contains not less than 10 vibrational levels. For attractive states, the equilibrium distances, $R_0$,  are presented in Table \ref{table:1}, with the corresponding energy minima.

The lifetime of   in an excited electronic state, about  $10^{-9}-10^{-7}$ s, is much greater than the duration of a collision. Being formed in an excited repulsive state,  $H_2^+$ dissociates immediately to proton and  hydrogen atom (direct channel to  produce $H$ in an excited state), but if $H_2^+$  is formed in an excited attractive state, there is a possibility to descend to a lower-lying quasi-molecular state (repulsive or attractive) with a subsequent dissociation or cascade down. This effect constitutes an indirect channel of producing  $H$  in an excited state.

In our treatment, we assume that $H_2^+$  are created at a large distance between protons in  excited  $\sigma$ electronic states (thick arrow in Fig.1). There are five repulsive states -- $5s, 6d, 8h$  and $7f, 6p$  (not shown in Fig 1) -- that correlate with states of the hydrogen atom with $n=5$   at  $R=\infty$. In these states $H_2^+$   rapidly dissociates into  hydrogen atom and proton. As for attractive states, being in states $7g$  and $8i$ the quasi-molecules rapidly  relax to the lowest vibrational level and then descend to lower-lying states according to the Franck-Condon principle (vertical transitions). Transitions from remaining attractive states $9j,10l$   and $9k$ are inhibited  by the extremely small Franck-Condon factors (their minima are located too far from the minima of lower-lying attractive states).

Taking into account that dipole transitions are allowed only between states of opposite parity, the problem reduces to the treatment of the following transitions
\begin{equation}
\left. \begin{array}{l}
7g\\
8i
\end{array} \right\} \to \left\{ {6f,\;5p,\;5f,\;4p,\;3p,\;2p\;} \right.
\label{eq_E5}
\end{equation}
for the direct channel and
\begin{equation}
\left. \begin{array}{l}
7g\\
8i
\end{array} \right\} \to \left\{ \begin{array}{l}
7h \to \;7i \to \left\{ {5f,\;4p,\;3p,\;2p,\;} \right.\\
7h \to \;7i \to 6h \to \;5g \to \left\{ {3p,\;2p,} \right.\\
7h \to \;7i \to 6h \to 5g \to \;4f \to \;3d \to \left\{ {2p,\;} \right.\\
7h \to \left\{ {3s,\;2s} \right.\\
7h \to 7i \to 6h \to \left\{ {2s} \right.
\end{array} \right.
\label{eq_E6}
\end{equation}
for the indirect channel.

\subsection{Formation of the hydrogen atom }
The total probability of the various processes is a sum of the probabilities of the separate processes just as the duration of consecutive processes is a sum of the separate periods. Taking that effect into account and following equation  (\ref{eq_E5}), one can write for the probability per unit time of a direct formation of atomic hydrogen in the ground state from quasi-molecular states $7g$  and $8i$  that
\[W_{dir}=\Omega (7g){{W}_{dir}}(7g)+\Omega (8i){{W}_{dir}}(8i),\]
in which
\begin{equation}
\begin{array}{l}
{W_{dir}}(i) = \rho (i \to 2p) + {\left[ {{\rho ^{ - 1}}(i \to 6f) + W_\infty ^{ - 1}(6f)} \right]^{ - 1}}\\
 + {\left[ {{\rho ^{ - 1}}(i \to 5p) + W_\infty ^{ - 1}(5p)} \right]^{ - 1}} + {\left[ {{\rho ^{ - 1}}(i \to 5f) + W_\infty ^{ - 1}(5f)} \right]^{ - 1}}\\
 + {\left[ {{\rho ^{ - 1}}(i \to 4p) + W_\infty ^{ - 1}(4p)} \right]^{ - 1}} + {\left[ {{\rho ^{ - 1}}(i \to 3p) + W_\infty ^{ - 1}(3p)} \right]^{ - 1}}.
\end{array}
\label{eq_E7}
\end{equation}
Here $\Omega(i)$  is the probability that $H_2^+$  is created in the state $|i\rangle$, $\rho (i \to j)$   is the probability per unit time of transition from  $|i\rangle$ to $|j\rangle$ quasi-molecular state; $W_{\infty}$ is the probability per unit time of a cascade downward to the ground state of  $H$ after dissociation of $H_2^+$.

Following equation (\ref{eq_E6}), one can write for the probability of an indirect formation of atomic hydrogen in the ground state from quasimolecular states  $7g$ and  $8i$ that
\[W_{indir} = \Omega (7g){W_{indir}}(7g) + \Omega (8i){W_{indir}}(8i)\]
in which
\begin{equation}\label{eq_E8}
\begin{array}{l}
{W_{indir}}(i) = \left[ {{\rho ^{ - 1}}(i \to 7h) + {\rho ^{ - 1}}(7h \to 7i)} \right.\\
{\left. { + {\rho ^{ - 1}}(7i \to 2p)} \right]^{ - 1}} + \left[ {{\rho ^{ - 1}}(i \to 7h) + {\rho ^{ - 1}}(7h \to 7i)} \right.\\
{\left. { + {\rho ^{ - 1}}(7i \to 6h) + {\rho ^{ - 1}}(6h \to 5g) + {\rho ^{ - 1}}(5g \to 2p)} \right]^{ - 1}}\\+
\left[ {{\rho ^{ - 1}}(i \to 7h) + {\rho ^{ - 1}}(7h \to 7i) + {\rho ^{ - 1}}(7i \to 6h)} \right.\\
 + {\rho ^{ - 1}}(6h \to 5g) + {\rho ^{ - 1}}(5g \to 4f) + {\rho ^{ - 1}}(4f \to 3d)\\
{\left. { + {\rho ^{ - 1}}(3d \to 2p)} \right]^{ - 1}}.
\end{array}
\end{equation}
A sum of (\ref{eq_E7}) and (\ref{eq_E8}) defines the complete probability of the formation of atomic hydrogen in the ground state. In the above equations non-adiabatic transitions between quasi-molecular states are entirely ignored.
\begin{table}
\caption{Equilibrium distances, $R_0$ and energies of the lowest sixteen
                          electronic states of $H_2^ +$ at $R=R_0$; parabolic quantum numbers $n_1, n_2$ and parity of state are shown within parentheses.}
\begin{tabular}{|c|l|l|c|l|l|}
\hline
State         & $R_0,a_0$ & $U(R_0), \varepsilon_0$ & State         & $R_0,a_0$ & $U(R_0), \varepsilon_0$ \\ \hline\hline
$1s, (00_g)$ & 2.00      & -0.6026                 & $2p, (00_u)$  & 12.546    & -0.500061             \\
$3d, (01_g)$  & 8.83      & -0.1750                 & $4f, (01_u)$  & 20.92     & -0.1307                 \\
$5g, (02_g)$  & 23.90     & -0.0782                 & $6h, (02_u)$  & 40.52     & -0.0606                 \\
$4d, (11_g)$  & 17.85     & -0.0588                 & $7i, (03_g)$    & 47.36     & -0.0436                 \\
$6g, (12_g)$  & 33.64     & -0.0379                 & $8j, (03_u)$  & 68.17     & -0.0352                 \\
$7h, (12_u)$  & 56.09     & -0.0326                 & $9k, (04_g)$  & 79.23     & -0.0276                 \\
$8i, (13_g)$  & 59.68     & -0.0255                 & $10l, (04_u)$ & 103.94    & -0.0230                 \\
$9j, (13_u)$   & 84.55     & -0.0218                 & $7g, (22_g)$  & 49.31     & -0.0207                 \\ \hline
\end{tabular}
\label{table:1}
\end{table}
\subsection{Formation of $H_2^+$  in the ground state}
Repulsive energy term $2p$  of $H_2^+$  has a minimum at $R=12.546 a_0$  \citep{Landau}. This minimum, which is due to van der Waals forces, much shallower than that of ground-state term $1s$  (see Table \ref{table:1}). Adjusting an harmonic-oscillator potential to the numerical data, one can find that the potential well contains one vibrational level. Adjusting the Morse potential \citep{Morse} to the numerical data leads, notably, to the same result. Hence, hereafter $2p$  might be considered  an attractive state.

The existence of a bound state with  equilibrium distance near $R_0=2.0 a_0$  leads to a possibility of the formation of $H_2^+$  in the ground state. The complete probability per unit time of a transition from states  $7g$  and $8i$   to the ground state of $H_2^+$  is
\[{W_{mol}} = \Omega (7g){W_{H_2^ + }}(7g) + \Omega (8i){W_{H_2^ + }}(8i)\]
in which
\begin{equation}\label{eq_E9}
\begin{array}{l}
{W_{H_2^ + }}(i) = \left\{ {\left[ {\rho (i \to 2p) + \left( {{\rho ^{ - 1}}(i \to 7h) + {\rho ^{ - 1}}(7h \to 7i)} \right.} \right.} \right.\\
{\left. { + {\rho ^{ - 1}}(7i \to 2p)} \right)^{ - 1}} + \left( {{\rho ^{ - 1}}(i \to 7h) + {\rho ^{ - 1}}(7h \to 7i)} \right.\\
{\left. { + {\rho ^{ - 1}}(7i \to 6h) + {\rho ^{ - 1}}(6h \to 5g) + {\rho ^{ - 1}}(5g \to 2p)} \right)^{ - 1}}\\
 + \left( {{\rho ^{ - 1}}(i \to 7h) + {\rho ^{ - 1}}(7h \to 7i) + } \right.{\rho ^{ - 1}}(7i \to 6h)\\
 + {\rho ^{ - 1}}(6h \to 5g) + {\rho ^{ - 1}}(5g \to 4f) + {\rho ^{ - 1}}(4f \to 3d)\\
{\left. {{{\left. {{{\left. { + {\rho ^{ - 1}}(3d \to 2p)} \right)}^{ - 1}}} \right]}^{ - 1}} + {\rho ^{ - 1}}(2p \to 1s)} \right\}^{ - 1}}.
\end{array}
\end{equation}

 There is thus  an additional channel -- a molecular channel that leads to the formation of $H_2^+$   in its ground state. This statement becomes obvious when one takes into account that term  $2p$   of $H_2^+$   with $m=1$   has a deep minimum at $R=7.93{a_0}$  \citep{BR1968}. The probability of a transition from this state  to the ground state is hence substantial. An estimate of the contribution of  $\pi$ terms in the formation of  $H_2^+$  in the ground state is a separate task, to be treated in forthcoming work.

 \subsection{Influence on the ionization energy }
 We seek to show how a participation of a second proton in a recombination alters the binding energy of an electron. According to equation  (\ref {eq_E1}), one can write for the electron energy at large $R$  that
 \begin{equation}\label{eq_F10}
 {\varepsilon _{{n_1}{n_2}\left| m \right|}}(R) = {\varepsilon _{{n_1}{n_2}\left| m \right|}}(\infty ) + \frac{{3n({n_1} - {n_2})}}{{2{R^2}}},
\end{equation}
in which  ${\varepsilon _{{n_1}{n_2}\left| m \right|}}(\infty ) =  - 1/2{n^2}$ is the electron energy in the isolated hydrogen atom.

Inserting in (\ref {eq_F10}) the average distance between protons during the pre-recombination period of evolution of the universe, which might be assumed to be  $\bar R = 2{r_n}$ \citep{KerMNRAS} in which $r_n=2{n^2}$  is the radius of the hydrogen atom in the excited state, we obtain that
\begin{equation}\label{eq_F11}
{\varepsilon _{{n_1}{n_2}\left| m \right|}}(\bar R) = {\varepsilon _{{n_1}{n_2}\left| m \right|}}(\infty ) + \frac{{3({n_1} - {n_2})}}{{32{n^3}}}.
\end{equation}

Equation (\ref {eq_F11}) shows that the participation of a second proton in the process increases the binding energy of an electron if  $n_2>n_1$, and it decreases the  binding energy if $n_1>n_2$. For excited states with $n\gg1$  the binding energy attains a maximal value  $19/32n^2$  when $n_1=m=0$. We thus obtain that, in the perturbed hydrogen atom, a maximal deviation of the ionization energy from its value in  unperturbed $H$ can attain 18.75\%.
\section{Results of calculations}
We proceed to calculate the probabilities involved in equations (\ref{eq_E7})-(\ref{eq_E9}). The probability of bound-bound and bound-free radiative transitions in $H_2^+$  is defined as \citep{Heit}
\begin{equation}\label{eq_E10}
\rho (i \to f) = \frac{{4w_{if}^3}}{{3{c^3}}}{\left| {{d_{if}}} \right|^2}.
\end{equation}
Here  $w_{if}$ is the frequency of an emitted photon, $c$  is speed of light, and $d_{if}$  is the transition matrix element defined with wavefunctions of $H_2^+$.

In the adiabatic approximation the wavefunctions of $H_2^+$  are representable as a product of two functions $\Psi  = \chi \psi $, in which $\psi (\vec r,R)$  and  $\chi(\vec R)$ describe motion of an electron and protons, respectively. Inserting  $\Psi  = \chi \psi $ into the transition matrix element and taking into account that   $\psi $ depends smoothly on $R$, we obtain that
\begin{equation}\label{eq_E11}
{d_{if}}(R) = \left\langle {{\chi _f}} \right|\left. {{\chi _i}} \right\rangle \left\langle {{\psi _f}} \right|z\left| {{\psi _i}} \right\rangle.
\end{equation}
In (\ref{eq_E11}) $\left\langle {{\psi _f}} \right|z\left| {{\psi _i}} \right\rangle$  is the matrix element of the  electric dipole moment; $ \left\langle {{\chi _f}} \right|\left. {{\chi _i}} \right\rangle$  is the vibrational overlap integral or the Franck-Condon factor. For a transition from a bound to anti-bound state $\chi_f$  should be replaced in $ \left\langle {{\chi _f}} \right|\left. {{\chi _i}} \right\rangle$  with the appropriate wavefunction  $\Phi_f$ describing the nuclear motion in a repulsive field. Explicit expressions for the Franck-Condon factor and overlap integral $ \left\langle {{\Phi _f}} \right|\left. {{\chi _i}} \right\rangle$ are presented in appendix A.

\begin{table}
\caption{Matrix elements of electric dipole strength at  $R=R_0$ involving the lowest ten  electronic states of $H_2^ +$; $R_0$ is the equilibrium distance of the upper state}
\begin{tabular}{|l|r|l|r|}
  \hline
  Transition & $\left\langle f \right|z\left| i \right\rangle ,{a_0}$ & Transition & $\left\langle f \right|z\left| i \right\rangle ,{a_0}$ \\
  \hline\hline
  $7g \to 2p$ & -0.003 & $8i \to 2p$ & 0.048 \\
  $7g \to 7h$ & 3.031 & $8i \to 7h$ & 10.589 \\
  $7h \to 7i$ & -1.130 & $7i \to 6h$ & 6.503 \\
  $6h \to 5g$ & -9.673 & $5g \to 4f$ & 3.004 \\
  $4f \to 3d$ & 6.924 & $5g \to 2p$ & -0.287 \\
  $3d \to 2p$ & 0.795 & $2p \to 1s$ & -6.246 \\
  \hline
\end{tabular}
\label{table: 2}
\end{table}

Our purpose is to calculate the ratio
\begin{equation}\label{eq_E12}
\eta (i) = \frac{{{W_{H_2^ + }}(i)}}{{{W_{dir}}(i) + {W_{indir}}(i)}}
\end{equation}
for $i=3d,5g, 7i,8i,7g$. This ratio does not depend on $\Omega(i)$ and allows us to estimate relative contribution of formation of $H_2^+$  in its ground state in recombination.

In $H_2^+$ the transition probabilities between two attractive states are values of order ${10}^{-10}-{10}^{-14}$ (per atomic unit of time), whereas the transition probabilities from an attractive to  repulsive states $6f, 5p, 5f, 4p$ and $3p$ are much smaller. Neglecting small terms, we thereby simplify  $\eta(i)$. The appropriate expressions are presented in appendix B.

Matrix elements of the electric dipole strength corresponding to the transitions involved in equations (\ref{eq_Bnew1})-(\ref{eq_Bnew4}) are collected in Table \ref{table: 2}. For the lowest four states the data are taken from  \cite{RamPeek}; for the highly excited states matrix elements are calculated with an algorithm developed by \cite{Dev2005}, and  employing asymptotic wavefunctions for  $H_2^+$ \citep{Ker2003}.

Using equations (\ref {eq_Bnew1})-(\ref {eq_Bnew4}), we calculated $\eta(i)$ and  obtained that  $\eta (3d) = 2.5 \times {10^{ - 5}}$, $\eta (5g) = 5.1 \times {10^{ - 1}}$, $\eta (7i) = 5.3 \times {10^{ - 1}}$,  and  $\eta (7g)=\eta (8i) = 5.4 \times {10^{ - 1}}$. This quantitative analysis thus reveals that, apart from $\eta(3d)$, that is a small value, all other $\eta(i)$  are values of order unity and are nearly equal. These obtained results clearly show that, in the pre-recombination period of evolution of the universe, the formation of  $H_2^+$ in the ground state introduced an important contribution, together with formation of  $H(1S)$, to the recombination.

\section{Conclusions}
In the present work, we have analysed quantitatively the recombination of an electron and a proton when the nearest neighbouring proton  participates in the process. The system of colliding particles is considered a quasi-molecule,  $H_2^+$, temporarily formed during a collision. This analysis has been implemented in an adiabatic approximation, in which the lowest thirty electronic states of $H_2^+$  with  $m=0$ ($\sigma$ states) were involved. The presence of another proton reduces the symmetry of a field experienced by an electron from spherical to axial. This reduction of symmetry leads in turn to the radiative transitions that are forbidden in the recombination of an electron on an isolated proton.

In the developed scheme of calculations our inclusion of higher electronic states leads to no qualitatively new and formidable problem -- it only complicates the treatment. We hence expect that $\eta(i)$  calculated for higher quasi-molecular states will be near  $\eta(i)$ obtained for states of the present large number, even though finite. Our expectation is based on the fact that the Franck-Condon factors decrease rapidly for highly excited attractive states (the locations of the energy minima are shifted toward large internuclear distances); accordingly, the participation of highly excited quasi-molecular states in the recombination declines.

The main results obtained in this work are that the QMR allows formation of $H_2^+$  in its ground state and that the probability of this process is comparable with the probability of recombination of an electron on an isolated proton. Another important result is that the QMR increases the binding energy of an electron during the recombination period. The participation of a second proton in the recombination thus increases the binding energy of an electron involved in the process and decreases the rate of recombination of hydrogen. An inspection of $\eta(i)$ shows that an inclusion of a molecular channel in the recombination maintains an unchanged rate of disappearance of free electrons, but accelerates the loss of free protons about 1.5 times.

In the pre-recombination period of evolution of the universe, the primordial plasma was thus composed of neutral hydrogen and helium atoms, hydrogen molecular ions $H_2^+$, protons and electrons, all exposed to the radiation field. A significant number of formed $H_2^+$  were obviously dissociated through photo-excitation in the repulsive quasi-molecular state or through a collision with other particles \citep{Coppola2011, Galli2013}. An evaluation of these processes and their influence on the rate of disappearance of  $H_2^+$ is a separate problem.

The quantitative analysis that we have performed confirms that the QMR plays an important role, and, accordingly, must be taken into account for a complete treatment of the cosmological recombination. As a possible significant outcome, we note that inclusion of the quasi-molecular corrections in the cosmological recombination can increase the Hubble constant estimated from analysis of the cosmic microwave background data and, accordingly, decrease the tension with local measurements  \citep{Beradze2019}.

The next step in the solution of the problem of cosmological recombination is a calculation of matrix elements for the initial free-bound transitions using the two-Coulomb-centre wavefunctions derived for the continuous spectrum \citep{Ker2019}. A knowledge of these data allows us to determine the absolute values of probabilities of the formation of $H$ and $H_2^+$   in the pre-recombination stage of evolution of the universe.

\section*{Acknowledgements}

One of us (TK) thanks Dr M. Gogberashvili for helpful discussions of the Hubble tension problem.







\appendix

\section{}
For two harmonic oscillators with disparate both equilibrium position and vibrational frequency the Franck-Condon factor is expressible as  \citep{Ch}
\begin{equation}\label{eq_B1}
\begin{array}{l}
\left\langle {{\chi _\nu }} \right|\left. {{\chi _{{\nu'}}}} \right\rangle  = {\left( {\frac{{A{e^{ - s}}}}{{{2^{\nu  + {\nu'}}}\nu !{\nu'}!}}} \right)^{1/2}}\sum\limits_{k = 0}^\nu  {\sum\limits_{k' = 0}^{\nu'}  \binom{\nu}{k}\binom{\nu'}{k'}{H_{\nu  - k}}(b)}  \\
{H_{{\nu'} - {k'}}}({b'}){\left( {2\sqrt \alpha  } \right)^k}{\left( {2\sqrt {{\alpha '}} } \right)^{{k'}}}I(K),
\end{array}
\end{equation}
in which Hermite polynomial $H_\nu(x)$  corresponds to vibrational state $\chi _\nu$, $A = 2\sqrt {\alpha {\alpha'}} /(\alpha  + {\alpha'})$, $s = \alpha {\alpha'}{d^2}/(\alpha  + {\alpha'})$, $b =  - {\alpha'}\sqrt \alpha  d/(\alpha  + {\alpha'})$, ${b'} = \alpha \sqrt {{\alpha'}} d/(\alpha  + {\alpha'})$ in which $\alpha=\omega/\hbar$, $\alpha'=\omega'/\hbar$, $d$ is the displacement between the two oscillators and $\omega$  is the angular frequency of the oscillator; $I(K)=0$ for $k+k'$ odd; $I(K)=(2K-1)!!/(\alpha+\alpha')^K$ for $k+k'$ even.

The wavefunction describing nuclear motion in a repulsive field that is defined with the first two terms in equation (\ref{eq_E1}) reads

\begin{equation}\label{eq_B2}
{\Phi _f}(R) = {C_\kappa }{(\kappa R)^{ - 1/2}}{J_{\sqrt {\gamma  + 1/4} }}(\kappa R)
\end{equation}
in which $J_{\sqrt {\gamma  + 1/4} }(\kappa R)$  is a Bessel function of the first kind, ${\kappa ^2} = 2\mu \left( {{U_i}({R_0}) + 1/(2{n^2})} \right)$, $\gamma  = 3\mu n({n_1} - {n_2})$,  $\mu$ is the reduced mass of two protons and $C_\kappa$  is a normalizing factor. A nuclear rotational motion is ignored in the  derivation of (\ref{eq_B2}).

 For a transition from an attractive electronic state with vibrational quantum number $\nu=0$  to a repulsive state, the overlap integral is defined as
\begin{equation}\label{eq_B3}
\left\langle {{\Phi _f}} \right|\left. {{\chi _i}} \right\rangle  = \int\limits_0^\infty  {{\Phi _f}(R){\chi _0}(R)} dR ,
\end{equation}
in which
\begin{equation}\label{eq_B4}
{\chi _0}(R) = {C_0}{e^{ - \frac{{\alpha \mu {{(R - {R_0})}^2}}}{2}}}
\end{equation}
and $C_0=(\alpha\mu/\pi)^{1/4}$ is the normalizing factor.

\section{}
Here are presented the simplified expressions:

\noindent for $\eta (3d)$
\begin{equation}\label{eq_Bnew1} \indent
         \eta (3d) = \rho (2p \to 1s){\left\{ {\rho (3d \to 2p) + \rho (2p \to 1s)} \right\}^{ - 1}},
\end{equation}

\noindent for $\eta (5g)$
\begin{equation}\label{eq_Bnew2} \indent
\begin{array}{l}
\eta (5g) = A(5g)\rho (2p \to 1s)\left\{ {\rho (5g \to 4f)} \right.\\
 \cdot \rho (4f \to 3d)\rho (3d \to 2p) + A(5g)\\
 \cdot {\left. {\left[ {\rho (5g \to 2p) + \rho (2p \to 1s)} \right]} \right\}^{ - 1}},
\end{array}
\end{equation}
in which
\[\begin{array}{l}
A(5g) = \rho (5g \to 4f)\rho (4f \to 3d)\\
 + \left[ {\rho (5g \to 4f) + \rho (4f \to 3d)} \right]\rho (3d \to 2p),
\end{array}\]

\noindent for $\eta (7i)$
\begin{equation}\label{eq_Bnew3} \indent
\begin{array}{l}
\eta (7i) = A(7i)B(7i)\rho (2p \to 1s)\left\{ {\rho (7i \to 6h)} \right.\\
 \cdot \rho (6h \to 5g)\left[ {A(7i)\rho (5g \to 4f)\rho (4f \to 3d)} \right.\\
\left. { \cdot \rho (3d \to 2p) + B(7i)\rho (5g \to 2p)} \right] + A(7i)B(7i)\\
{\left. { \cdot \rho (2p \to 1s)} \right\}^{ - 1}},
\end{array}
\end{equation}
in which
\[
\begin{array}{l}
A(7i) = \rho (7i \to 6h)\rho (6h \to 5g) + \rho (5g \to 2p)\\
 \cdot \left[ {\rho (7i \to 6h) + \rho (6h \to 5g)} \right],
\end{array}
\]
\[ \begin{array}{l}
B(7i) = \rho (7i \to 6h)\rho (6h \to 5g)\rho (5g \to 4f)\\
 \cdot \rho (4f \to 3d) + \rho (3d \to 2p)\left[ {\rho (6h \to 5g)} \right.\\
 \cdot \rho (5g \to 4f)\rho (4f \to 3d) + \rho (7i \to 6h)\\
 \cdot \rho (5g \to 4f)\rho (4f \to 3d) + \rho (7i \to 6h)\\
 \cdot \rho (6h \to 5g)\rho (4f \to 3d) + \rho (7i \to 6h)\\
 \cdot \left. {\rho (6h \to 5g)\rho (5g \to 4f)} \right].
\end{array}
\]
and for $\eta(i)$ ($i=7g, i=8i$)
\begin{equation}\label{eq_Bnew4} \indent
\begin{array}{l}
\eta (i) = A'(i)B'(i)C'(i)\rho (2p \to 1s)\left\{ {A(i)B'(i)C'(i)} \right.\\
 + A'(i)B(i)C'(i) + A'(i)B'(i)C(i)\\
 + {\left. {A'(i)B'(i)C'(i)\rho (2p \to 1s)} \right\}^{ - 1}},
\end{array}
\end{equation}
in which
\[\begin{array}{l}
A(i) = \rho (i \to 7h)\rho (7h \to 7i)\rho (7i \to 2p),\\
\\
B(i) = \rho (i \to 7h)\rho (7h \to 7i)\rho (7i \to 6h)\\
 \cdot \rho (6h \to 5g)\rho (5g \to 2p),\\
 \\
C(i) = \rho (7g \to 7h)\rho (7h \to 7i)\rho (7i \to 6h)\rho (6h \to 5g)\\
 \cdot \rho (5g \to 4f)\rho (4f \to 3d)\rho (3d \to 2p),\\
 \\
 A'(i) = \rho (i \to 7h)\rho (7h \to 7i) + \left[ {\rho (i \to 7h)} \right.\\
\left. { + \rho (7h \to 7i)} \right]\rho (7i \to 2p),\\
\\
B'(i) = \rho (i \to 7h)\rho (7h \to 7i)\rho (7i \to 6h)\rho (6h \to 5g)\\
 + \left[ {\rho (i \to 7h)\rho (7h \to 7i)\rho (7i \to 6h) + \rho (i \to 7h)} \right.\rho (7h \to 7i)\\
 \cdot \rho (6h \to 5g) + \rho (i \to 7h)\rho (7i \to 6h)\rho (6h \to 5g)\\
\left. { + \rho (7h \to 7i)\rho (7i \to 6h)\rho (6h \to 5g)} \right]\rho (5g \to 2p).
\end{array}
\]
\[\begin{array}{l}
{C'} = \rho (7g \to 7h)\rho (7h \to 7i)\rho (7i \to 6h)\rho (6h \to 5g)\\
 \cdot \rho (5g \to 4f)\left[ {\rho (4f \to 3d) + \rho (3d \to 2p)} \right] + \rho (7g \to 7h)\\
 \cdot \rho (7h \to 7i)\rho (7i \to 6h)\rho (4f \to 3d)\left[ {\rho (6h \to 5g) + \rho (5g \to 4f)} \right]\\
 \cdot \rho (3d \to 2p) + \rho (7g \to 7h)\left[ {\rho (7h \to 7i) + \rho (7i \to 6h)} \right]\rho (6h \to 5g)\\
 \cdot \rho (5g \to 4f)\rho (4f \to 3d)\rho (3d \to 2p) + \rho (7h \to 7i)\rho (7i \to 6h)\\
 \cdot \rho (6h \to 5g)\rho (5g \to 4f)\rho (4f \to 3d)\rho (3d \to 2p).
\end{array} \]


\bsp	
\label{lastpage}
\end{document}